# Unveiling the Property of Radiation: Angular Synthetic Aperture for Omnidirectional Fourier Imaging System with Centimeter Field of View

**JINXI HUANG**[*]

*Department of Electrical and Computer Engineering, University of California, Santa Barbara, Santa Barbara, CA, 93106, USA*
*\*Corresponding author: jinxihuang@umail.ucsb.edu*

**We demonstrate a Fourier Imaging system with centimeter field of view(FOV), and 0.0042°angular resolution at C band, for the entire 180° radiation angular range. To pursue omnidirectional Fourier imaging without aberration, a technique called angular synthetic aperture is employed, by multi-perspective imaging and stitching. The effect of tilting-introduced virtual lens is pointed out and dealt with in our system. For the sake of limited sensor size and aberration spot size, Fourier imaging at each perspective is restricted to angle range smaller than 3°. The imaging system with a 4f demagnification component is prudently designed to magnify Fourier imaging, relieving the pressure of pixel size of sensor for high sampling rate, meanwhile guarantee the 1cm object FOV within ±1.5° range. All the system is automatically controlled, which is crucial for applications such as online-optimization for optical phased array(OPA). As far as we know, this is the state-of-art omnidirectional Fourier imaging system as to field of view and resolution.**

    Fourier imaging keeps as a vital instrumentation for tons of applications, including radiation and scattering characterization for OPA[1], metamaterial[2], LED and grating ruler. In addition, some modern holography-based microscopy, for example optical ptychography[3], is in need of Fourier imaging. An ideal lens can do Fourier transformation on its focal plane, which is in coincident with angular distribution of far field diffraction. However, due to finite aperture size, some Fourier component are damped; that at the extreme large angle even fully vanishes, since the lens can't collect incident planar wave scattered or radiated from the object with those angles. Intuitively, to collect all radiation and scattering from the object, namely broaden the field of view for both object field and Fourier field, lens system with large numerical aperture(NA) seems like a good candidate. Nevertheless, it introduces two terrible side effects. First, the strong aberration of large NA system completely blurs the Fourier image, which is a non-inversible and shift-variant low pass filter. It's a complex industrial issue to justify all aberrations (like spherical, coma, curvature) simultaneously of large NA system. Moreover, the clumsy and expensive system, which may well even call for 10cm aperture lens, immersed imaging and sensor shifting intensively challenges cost and operation technique. Having developed successfully in holography, people use illumination from different angle[4] and structured-illumination[5] to introduce a linear spatial phase shift, so as to shift the high spatial frequency component into the pass band of Fourier imaging system. This brilliant technology even shines a point on beyond-diffraction-limit imaging. Unfortunately, it's limited to applications of passive illuminated imaging, and the imaging for Fourier component far beyond diffraction limit is increasingly harder as aberration still plays the role of blurring. Last but not least, the size of Fourier image is proportional to focal length. So here comes the dilemma between image size and sensor size: larger magnification benefits Fourier resolution as the pixel size (normally 10um level) limits sampling rate; while the sensor is comparatively small and fail to capture wide Fourier image range.

    Inspired by radioative telescope synthetic aperture, we develop the technology angular synthetic aperture(ASA), as shown in fig.1. In this system, lens tube T will rotate along the center of the object.

At a certain perspective, not all spatial frequency component but only that with angle near the perspective is needed to be imaged. By sweeping all perspectives and stitching the partial Fourier images together, We can reconstruct the omnidirectional far field image. Thanks to this distributed aperture, at each perspective it's not necessary to cover a wide angular range of spatial spectrum information and consequently aberration is weakened to negligible level with deliberate design of the optic system. This relieves the dilemma between sensor size and pixel size. Thanks to the stitching technology to equivalently enlarge sensor size, imaging system design just needs to focus on exploiting pixel size by broadening object FOV for high far field resolution, meanwhile magnifying the far field to match the system resolution and camera resolution. For passive illuminated holography superresolution, angular synthetic aperture successfully solves the problem of aberration for the component close to twice the diffraction limit in Fourier space. For characterization of radiative field, ASA is able to approach diffraction limit as well.

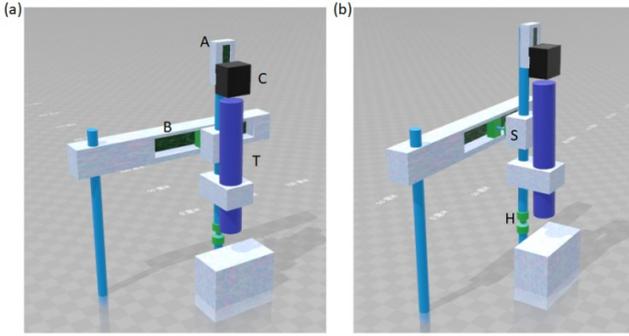

Fig. 1. The mechanic system contains the following parts: linear stage B, with a bearing fixed on the stage and connected to sleeve S which can slide on the fixing rod of the lens system. Once the stage slides, the bearing will rotate and the sleeve will slide, pushing the rod rotating along the hinge H underneath. This system transforms linear displacement to rotation. Camera C can shift along stage A to be perfectly located Fourier plane at every perspective.

However, when the object is tilted, the Fourier transformation function by lens should be modified correspondingly. Without tilting, two forms of expression can describe Fresnel diffraction between two parallel plane with distance d, , namely the filtration form and quadratic phase-modulated Fourier transformation (QPM-FT) form

$$D(x') = \int O(x)\exp(j2\pi\sqrt{d^2+(x'-x)^2}/\lambda)dx$$
$$\approx \begin{cases} O(x) \otimes \exp(j\pi x^2/\lambda d) & filtration \\ \exp(j\pi x'^2/\lambda d)\mathfrak{I}(O(x)\exp(j\pi x^2/\lambda d)) & QPM-FT \end{cases} \quad (1)$$

Here D(x') is diffracted field, O(x) is the object field. It's known that lens modulates a quadratic phase $\exp(-j\pi x'^2/\lambda f)$ to incident field. In diffraction from lens to back focal plane(BFP), with distance f, this quadratic phase cancels the phase modulation inside QPM-FT, giving rise to the Fourier transformation on BFP. In addition, if object is placed on the front focal plane(FFP), the diffraction filtration from FFP to lens plane cancels the first quadratic phase term in QPM-FT from lens plane to BFP. Consequently, BFP gives perfect FT to object on FFP.

However, when the object is tilted, the Fresnel diffraction from object to lens front plane is supposed to be modified as

$$D(x') = \int O(x)\exp(j2\pi\sqrt{(d+x\sin\theta)^2+(x\cos\theta-x')^2}/\lambda)dx \quad (2)$$
$$\approx \int [O(x)\exp(j2\pi x\sin\theta/\lambda)\exp(j\pi\sin^2\theta x^2/d\lambda)]\exp(j\pi(x\cos\theta-x')^2/\lambda d)dx$$
$$= \frac{1}{\cos\theta}\left[O(x/\cos\theta)\exp(j\pi\tan^2\theta x^2/d\lambda)\exp(j2\pi x\tan\theta/\lambda)\right] \otimes \exp(j\pi x^2/\lambda d)$$

Compared with non tilted object, it gives birth to three effects. Above all, in the third term in square bracket, the object is linearly phase tuned and thus shifts high spatial frequency component into the pass band of our imaging system, which satisfies our motivation of rotating for visibility of high spatial frequency. This is equivalent to angular illumination for passive superresolution. Secondly, the object is magnified by $\cos\theta$. This is apparent as well, since the projection of planar wave range $\Delta k$ from tilted object plane onto vertical plane magnifies it by $\cos\theta$ and weakens the intensity correspondingly. In addition, a quadratic phase modulation, in the second term in the bracket, appears, which is equivalent to a virtual concave lens with focal length $d/\tan^2\theta$. This magic can be interpreted as follows: the wave front is approximately spherically spread, while our diffracted plane is flat, calling for the quadratic phase compensation. It was pointed out by plenty of holography researches[6] as well. This virtual lens effect transforms the equivalent illumination to point illumination at S=d+ $d/\tan^2\theta$. By the property of lens Fourier imaging, the Fourier plane is now on the imaging plane of the point source, which is determined by Gauss imaging function

$$\frac{1}{S}+\frac{1}{F}=\frac{1}{f}$$

Moreover, the Fourier imaging is magnified by

$$M = \frac{S-F}{S+f} = \frac{S(S-2f)}{S^2-f^2} \quad (3)$$

If the object is placed on FFP, Fourier plane is $f\tan^2\theta$ from BFP, with M= $\cos 2\theta/\cos^2\theta(1+\sin^2\theta)$

In summary, the Fourier imaging F(u) is

$$F((u-f\tan\theta)\cos\theta/M) = \exp(j\pi u\tan\theta/\lambda)\mathfrak{I}(O(x))$$

acquiring on plane F, at perspective $\theta$.

Besides the limitation of lens system, pixel size of camera limits sampling rate and thus resolution. As to our application, namely characterization for OPA around 1550nm, the tiniest pixel size of commercial camera sensor is 10um; while ours is even 25um. Lens system with long effective focal length(EFL) is not a wise choice for larger Fourier image, as smaller NA sacrifices the capability to collect light. As a result, design to magnify Fourier image is necessary, in which 4f demagnification system is a good candidate. As shown in figure 2, lens 1, 2 and 3 are confocally located, with 2,3 generating a demagnified 4f system, and lens 1 being the Fourier lens. Apertures of these lenses are 2.5cm, with focal length 10cm, 3cm, 10cm. All lenses are aspherical in order to overcome spherical aberration. In our future work, every single lens will be substituted by lens set such as balsaming lens in the layout to further overcome aberration. 4f system can linearly magnify object, even if the object is not ideally on the focal point; additionally, it won't introduce too much coma. Fourier imaging is 0.0042° /pixel.(25um/10cm/3(arc)=0.0042°). At each perspective,

2.74° far field is captured since our camera sensor is 480*640. To analysis diaphragm, we map the former two lenses to the object space, as fig2(a) shows. Within incident angle smaller than 3°, lens system can capture 1.5cm object FOV; from diffraction limit calculation, this enables 0.0042°, which indicates that we exploit the pixel sampling rate. In our future work, 10um pixel IR camera will be applied, which satisfies the requirement of sampling rate of our lens system by Nyquist law . Aberration simulation within the angular range is simulated by Zemax, shown in fig2(b). As can be seen, rms keeps smaller than pixel size, thus low pass filter by MTF of aberration doesn't affect the imaging quality. The error of our linear stage A (fig 1) is 1um, leading to 0.001° angular error of the perspective with 5.6cm from the stage to the hinge, which is negligible.

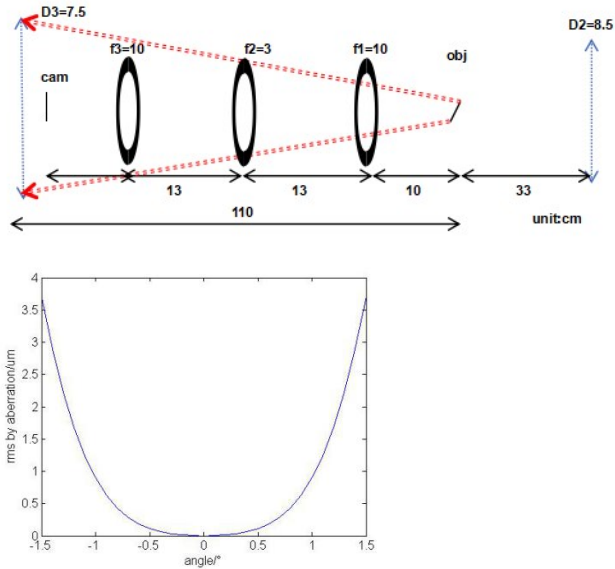

Fig. 2. (a)Fourier imaging system with 4f to magnify Fourier image, with diaphragms by each lens plotted with blue arrows. (b) aberration rms of the image system with different incident angle

Noticeably, two potential problems exist in this system. Firstly, the derivation of Fresnel diffraction is based on Huygens principle, assuming that each independent source uniformly illuminates on the receiving plane. However, for spherical-wave radiation source like dipole, with large incident angular range, this approximation is not very precise. Similarly, phase approximation in Fresnel diffraction is adoptable with small angular range as well. However, for Fourier imaging of radiation source with large aperture, it's highly possible that the radiation is relatively collimated, giving better precision of the approximation. The other problem is focus mismatch of two dimensions. Since our rotation is only in one dimension, the virtual lens only functions on that dimension, leading to focus mismatch between that dimension and the other. In our application to characterize one dimension waveguide grating and optimization for OPA, it's a negligible problem as we don't care about defocus of the other direction. However, applications like semiconductor light source characterization should be conducted on two dimensions. A potential solution is to add a movable cylinder lens inside the optical path to cancel the effect of virtual lens.

To verify the image system, we fabricates a group of weak waveguide gratings, whose periods are 520nm, 560nm, 580nm. The waveguides are on SOI platform with 1um buried oxide and 500nm top silicon. Rib waveguide is chosen, with 230nm rib depth and 800nm width, and consequently 3.1 effective index. We deposited 30nm thickness silicon nitride top and etched teeth as grating perturbation[7]. By wavelength sweeping from 1.5um to 1.6um, grating function

$$\sin\theta = n_{eff} - \lambda/\Lambda \qquad (4)$$

indicates the angular sweeping range shown in table below.

| period / Wavelength | 520nm | 560nm | 580nm |
|---|---|---|---|
| 1.6um | 1.36° | 13.88° | 19.72° |
| 1.5um | 14.61° | 27.15° | 33.22° |

To pursue weaker radiation, 0.8 duty cycle is picked up. Grating length is 1cm. By FDTD simulation, ratio of field intensity between input and output is plotted in fig3(a). As can be seen, it is close to 0.7. So for 1cm transmission, the effective aperture is roughly 1cm. Based on exponential decayed near field of grating, we can calculate the beam divergence from 0.0067 to 0.0092°. After intensity normalization at 1550nm with 520nm grating, beam intensity by both experimental imaging and simulation are plotted in fig3(b). As can be seen, their trend matches each other. Nevertheless, background noise and roughness scattering lift the peak value of beam, thus normalization makes the intensity more uniform and arises the error between experiment points and simulation curve. In fig4, the far field beam profiles at 1600nm(520nm period), 1560nm(520nm), 1520nm(520nm), 1580nm(560nm), 1540nm(560nm), 1550nm(580nm), 1500nm(580nm) are plotted. Our imaging system resolves the beam with 2 pixels, matching the theoretical calculation. Due to the limitation of shift range of stage B, and vertical space in lab, we only experimentally verify the imaging system with perspective angle smaller than 33°, as larger angle calls for longer traveling length of camera thanks to virtual lens effect.

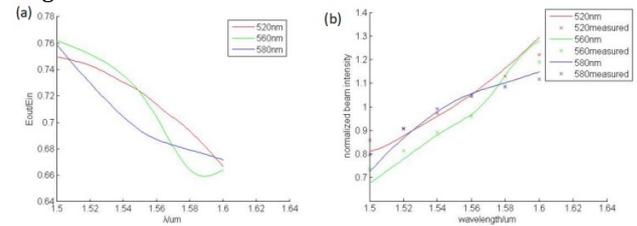

Fig. 3. (a)simulation of field intensity ratio between two ports. Three gratings with wavelength range 1.5-1.6um are shown(b)simulated normalized beam intensity and measured beam intensity

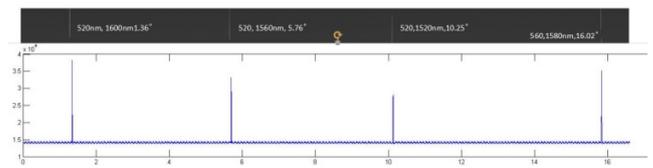

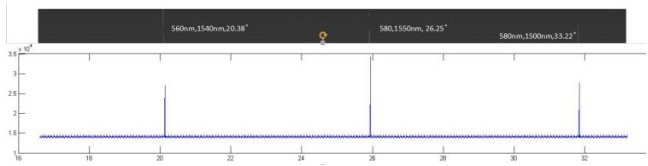

Fig. 4. panorama of beam measured within 33° for the beam with grating period, angle and wavelength shown inside, along with the beam plots

Adapting this system to passive illuminated imaging, we can add another reference channel to interfere with the Fourier image on the back focal plane to extract phase information. By employing integrated LiDAR as illumination, planar wave illumination can introduce lateral k to shift spatial frequency, which enables imaging twice beyond diffraction limit at most as fig 5 shows, which will be demonstrated in our future work.

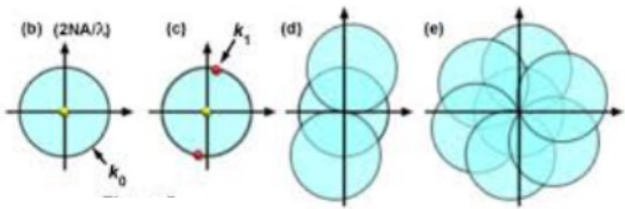

Fig. 5. This ASA Fourier imaging system broaden the pass band to omnidirectional, and integrated angular sweeping illumination source like integrated LiDAR shifts the component beyond diffraction limit into imaging system.

In this paper, we propose an angular synthetic aperture for omnidirectional Fourier imaging setup. With 2.5cm aperture of the 3 lens system, 1.5cm FOV is achieved with resolution 0.0042°, in one dimension. The solution to the defocus of the two dimensions due to perspective tilt by cylinder lens will be adopted in future work, which is crucial for two dimensional Fourier imaging. Besides the radiation characterization demonstrated in this paper, potential application to passive illuminated microscopy beyond diffraction limit is pointed out.

**Funding sources and acknowledgments.** Thanks Lockheed Martin Coherent Technology for discussion and advice. Thanks DARPA MTO MOABB for funding.